\documentclass{rsauthor_arxiv}

\providecommand{\tabularnewline}{\\}

\title{Modes of growth in dynamic systems}

\author{Timothy J. Garrett}
\address{Department of Atmospheric Sciences, University of Utah, Salt Lake City, Utah,  USA}
\usepackage{times}

\begin{document}
\maketitle

\begin{abstract}
RRegardless of a system's complexity or scale, its growth can be considered to be a spontaneous thermodynamic response to a local convergence of down-gradient material flows. Here it is shown how growth can be constrained to a few distinct modes that depend on the availability of material and energetic resources. These modes include a law of diminishing returns, logistic behavior and, if resources are expanding very rapidly, super-exponential growth. For a case where a system has a resolved sink as well as a source, growth and decay can be characterized in terms of a slightly modified form of the predator-prey equations commonly employed in ecology, where the perturbation formulation of these equations is equivalent to a damped simple harmonic oscillator. Thus, the framework presented here suggests a common theoretical under-pinning for emergent behaviors in the physical and life sciences. Specific examples are described for phenomena as seemingly dissimilar as the development of rain and the evolution of fish stocks.\end{abstract}

\section{Introduction}

Very generally, the physical universe can be considered as a locally
continuous distribution of energy and matter in the three dimensions
of space. Conservation laws dictate that total energy and matter
are conserved. The Second Law of Thermodynamics requires that a positive
direction for time is characterized by a net material flow from high
to low energy density. The rate of flow depends on the precise physical forces at hand. Spatial variability in flows allows for a local convergence in the density field.  \cite{Onsager1931_Part1,deGrootMazur1984}.

Most often though, we categorize our world in terms of discrete, identifiable
{}``things'', species, systems, or particles that require that we
artificially invoke some local discontinuity that distinguishes the
system of interest from its surroundings. Local variability within
the system is ignored, not necessarily because it doesn't exist, but
rather because we lack the ability or interest to resolve any finer
structure, at least in anything other than a purely statistical sense. The system
evolves according to flows to and from its surroundings as determined
by interactions across the predefined system boundaries. 

General formulations have been developed for characterizing rates
of potential energy dissipation within heterogeneous systems \cite{deGrootMazur1984,Kjelstrup2008}.
However, these do not explicitly express rates of growth for a discrete
system itself, nor how these system growth rates change with time.
What this paper explores is a unifying framework for expressing the
emergent growth of discrete systems, and discusses a few simple expressions
for the types of evolutionary phenomena that are thermodynamically possible.
Some, such as a law of diminishing returns, explosive or super-exponential growth, and
non-linear oscillatory behavior, have been identified in a very broad
range of scientific disciplines, ranging from cloud physics \cite{KorenFeingold2011}
to ecology \cite{Berryman1992} to energy economics \cite{Hook2010}.
These are shown here to have common physical roots.

\section{Growth and decay of flows}

\begin{figure*}
\noindent \begin{centering}
\includegraphics[width=124mm]{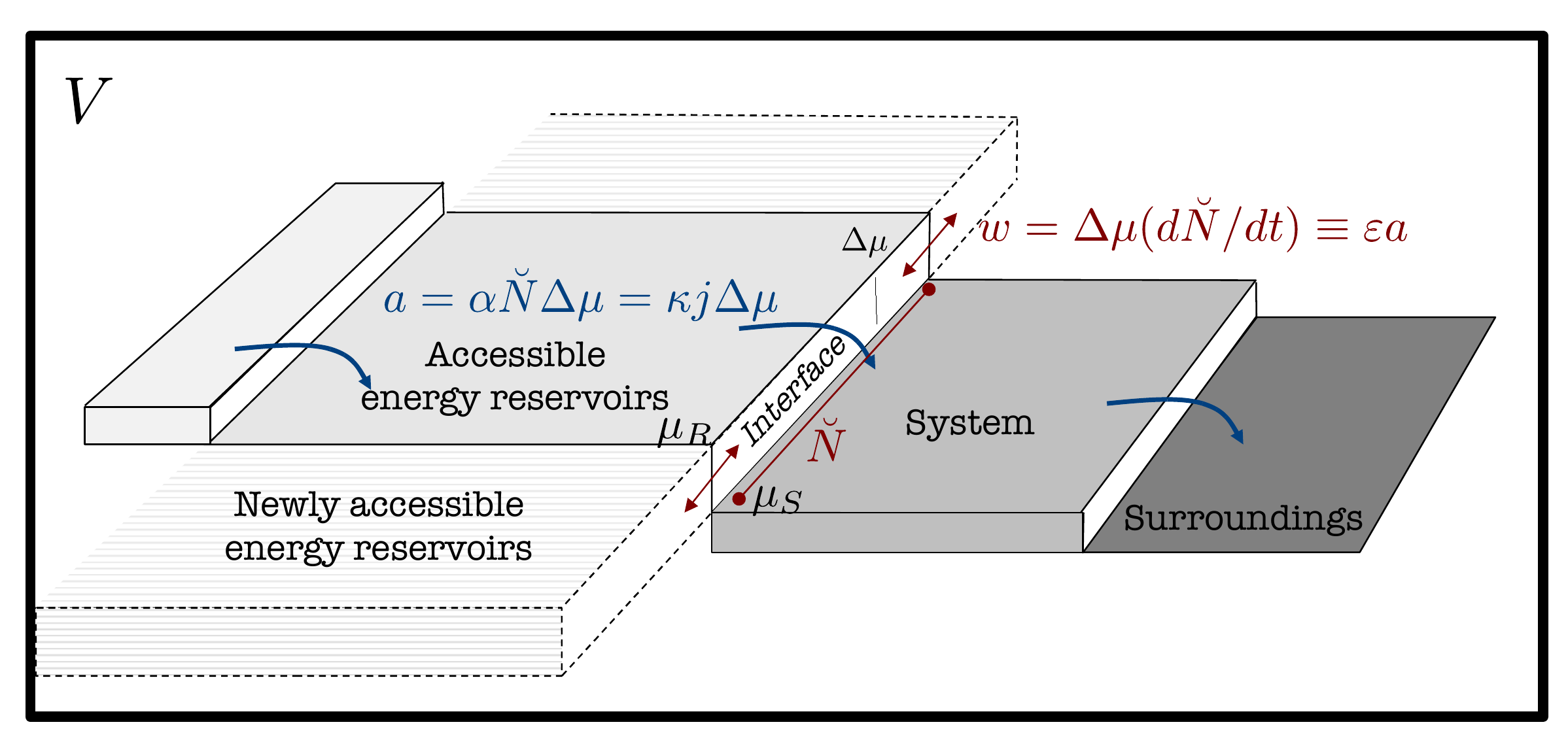}
\par\end{centering}

\caption{\emph{\label{fig:thermschematic}A schematic for the thermodynamic
evolution of a system at potential $\mu_{S}$ in response to flows
from a reservoir at higher potential $\mu_{R}$. The size of an interface
$\breve{N}$ along the gradient $\Delta\mu$ between the surfaces
determines the speed of downhill material flow $j$ and rate of energy
dissipation $a$. The interface itself grows or shrinks by doing work
at rate $w$, whose sign and efficiency $\varepsilon$ depends on
whether there is material convergence or divergence at $\mu_{S}$.
Growth expands flows by bootstrapping the interface into previously inaccessible
reservoirs. }}

\end{figure*}

Fig. \ref{fig:thermschematic} is an illustration of flows between discrete systems. A closed system of volume $V$ contains
a locally resolved fluctuation in pressure, density, or potential
energy per unit matter, that is represented as a discrete potential
{}``step'' $\mu_{S}$. Defined as a system at local thermodynamic
equilibrium, $\mu_{S}$ is resolved only as a surface of uniform potential
energy per unit matter. Thus, the nature of the step can be of arbitrary
internal complexity because, defined as a whole, the internal details
are unresolved. For example, $\mu_{S}$ might represent the sum of
the specific energies associated with any choice of force fields.
As an example, in the atmospheric sciences,  the mass specific moist static energy
$h_{m}$ of an air parcel is often employed as a simple, conserved
tracer, even if it is physically derived from the more complex sum of the potentials from gravitational forces, molecular motions, and molecular bonds. 

Because the step is itself open system, there are flows to and from it. Flows from some higher potential $\mu_{R}$ that characterizes a {}``reservoir' for the system are down
a small jump $\Delta\mu=\mu_{R}-\mu_{S}\ll\mu_{S}$ that separates the two steps.
There is an interface between the system and the reservoir that is defined by a
quantity of matter $\breve{N}$, such that the total magnitude of the
potential difference along the interface is $\Delta{G} = \breve{N}\Delta\mu$.
Because $\Delta\mu/\mu_{S}$ is small, the potential difference as defined is
never {}``far from equilibrium''  \cite{Nicolis2007}. 

While total matter and energy within the total volume $V$ are conserved,
a continuous flow redistributes potential energy downward.
This downhill dissipation of potential energy across $\Delta\mu$ manifests
itself as an energetic {}``heating'' of the system occupying $\mu_S$ at rate \begin{equation}
a=\alpha\breve{N}\Delta\mu\label{eq:a_alphanhat}\end{equation}
where $\alpha$ is a constant rate coefficient that can be related to the speed
of flow across the interface. The energetic heating is tied to a 
material flow $j$ through a coefficient $\kappa=\left(da/dj\right)/\Delta\mu$,
such that

\begin{equation}
a=\kappa j\Delta\mu\label{eq:a_jemu}\end{equation}
For example, matter falls down a gravitational potential gradient, and the
radiative dispersion of light can be expressed in terms of a flow of photons
from high to low energy density. The energetic and material convergence
into the fixed potential $\mu_{S}$ causes an orthogonal {}``stretching''
along $\mu_{S}$, allowing thermodynamic work to be done at rate \begin{equation}
w=\left(\frac{\partial\breve{N}}{\partial t}\right)_{\mu_{S},\mu_{R}}\Delta\mu\label{eq:work}\end{equation}
Work here is a linear expansion of the interface at constant density. Depending on whether or not there is net convergence
or divergence of flows at $\mu_S$, work can be either positive
or negative, in which case the interface $\breve{N}$ either grows or shrinks. From Eqs.
\ref{eq:a_alphanhat} and \ref{eq:work}, the dimensionless efficiency
$\varepsilon$ with which the dissipative heating $a$ is converted
to work $w$ is \begin{equation}
\varepsilon=\frac{w}{a}=\frac{1}{\alpha}\frac{d\ln\breve{N}}{dt}\label{eq:epsilon}\end{equation}
The sign of $\varepsilon$ dictates whether there is exponential growth
or decay in $\breve{N}$. Combining Eqs. \ref{eq:a_alphanhat} to
\ref{eq:epsilon} leads to \begin{equation}
\frac{d\ln j}{dt}=\frac{d\ln\breve{N}}{dt}=\alpha\varepsilon=\eta\label{eq:dlnadt}\end{equation}
where $\eta$ is the instantaneous rate at which flows into the system either grow or decay
(i.e. $j=j_{0}\exp\left(\eta t\right)$).

\section{Definition of the interface driving flows}

Supposing that the only resolved flows are those into the system from
a higher potential, it follows from Eq. \ref{eq:a_jemu} that the
material flow $j$ across the interface $\breve{N}$ results in an
increase in the amount of matter (or energy) in the system $N_{S}$
at the expense of the reservoir $N_{R}$ \begin{equation}
j=\alpha\breve{N}=\left(\frac{\partial N_{S}}{\partial t}\right)_{\mu_{S}}=-\left(\frac{\partial N_{R}}{\partial t}\right)_{\mu_{R}}\label{eq:dNc_dNe}\end{equation}
The relationship to energy dissipation is given by Eq. \ref{eq:a_jemu}.
Eq. \ref{eq:dNc_dNe} is proportional to an increase in the system
volume $V_{S}=N_{S}/n_{S}$, assuming no resolved internal variations
in the system density $n_{S}$. 

A first guess might be that the interface $\breve{N}$ is determined
by a product of concentrations $N_{S}N_{R}$. This is the approach that is most commonly
taken when modeling ecological populations \cite{Berryman1992} and
in the application of the logistic equation to long-range modeling
of national energy reserve consumption \cite{Bardi2009,Hook2010}.

However, perfect multiplication is only suitable when $N_{S}$ and
$N_{R}$ can be treated as being perfectly well-mixed. It is not possible to resolve flows between two components
of a perfect mixture. Rather, if $N_{S}$ and $N_{R}$ can be distinguished,
then they must interact through physical flows across some sort of
interface. Because fluid flows are always down a potential
gradient, the interface driving the flux $j$ from $N_{R}$
to $N_{S}$ is most appropriately defined as a concentration gradient normal to
a surface. It is the exterior surface of the system, and
a density gradient away from the surface, that provides the resolvable contrast allowing for a net
flow. 

Perhaps the simplest possible example of this physics is the diffusional growth of a particulate
sphere of radius $r$ within a larger volume $V$. Fick's Law dictates
that a concentration gradient $n$ drives a diffusive flux across
the sphere surface at rate \begin{equation}
j=\left.4\pi r^{2}\mathcal{D}n_{\mu_{S}}\frac{\partial\ln n}{\partial x}\right]_{x=r}\label{eq:dNldt_gradient}\end{equation}
where $\mathcal{D}$ is a diffusivity (units area per time) that expresses
the speed of material transfer across a surface with radius $r$ along
radial coordinate $x$. If the gradient is approximated as a small
discretized concentration jump between two potential surfaces $\Delta n=n_{\mu_{R}}-n_{\mu_{S}}$,
and the particle is small compared to the total volume, then the flux
of matter down the gradient is \begin{equation}
j\simeq4\pi r\mathcal{D}\Delta n=\frac{4\pi r\mathcal{D}}{V}N_{R}\label{eq:dNldt_jump-fixedV}\end{equation}
where $N_{R}=\Delta nV$ is the amount of matter in the higher potential
reservoir that is available to flow to the lower potential system
$N_{S}$. Note that it is a length dimension of $4\pi r$ that drives
flows rather than the whole particle volume or its surface area. 

In this respect, the electrostatic analogy for flows is that they
are proportional to a capacitance, which in cgs units has dimensions
of length. For shapes more complex than spheres \cite{WoodBakerCalhoun2001,Maia2005,Kooijman2010},
the length dimension can be retained but generalized such that the
flux equation given by Eq. \ref{eq:dNldt_jump-fixedV} becomes\begin{equation}
j=\lambda\mathcal{D}N_{R}\label{eq:dNs_jump}\end{equation}
where $\lambda$ is the effective length or capacitance density within
the the volume $V$. The flux of $N_{R}$ to $N_{S}$ has a time constant
$1/\left(\lambda\mathcal{D}\right)$. 

Interactions between particles or species are not always referenced
with respect to space. For example, thermal heating requires a radiation
pressure contrast, but the distance between the source and receiver
is not considered because light is so fast. Thus, a more convenient
expression for Eq. \ref{eq:dNs_jump} replaces the diffusivity with
a rate coefficient $\alpha$ that has dimensions of inverse time,
and replaces the length density or capacitance density $\lambda$
by $kN_{S}^{1/3}$ where $k$ is a dimensionless coefficient that
depends on the system geometry. The rate coefficient $\lambda\mathcal{D}$
is generalized to the geometry-independent expression $\alpha kN_{S}^{1/3}$
so that Eq. \ref{eq:dNs_jump} becomes

\begin{equation}
j=\alpha\breve{N}=\alpha kN_{S}^{1/3}N_{R}\label{eq:dNcdt_NcNe}\end{equation}
Thus, the material interface in Eq. \ref{eq:a_alphanhat} is proportional
to two quantities: the length density of the system within the total
volume, or its bulk to a one third power $N_{s}^{1/3}$, and the material
availability in the energy reservoir $N_{R}$. It is this product
that drives material flows at rate $j = \alpha\breve{N}$ and dissipates
energy at rate $a=\alpha\breve{N}\Delta\mu$. Flows are 
proportional to a surface area \emph{and }a
local gradient (e.g. $N_{S}^{1/3}$), rather than the system volume (e.g. $N_{S}$) or its surface area alone (e.g. $N_{S}^{2/3}$).

\section{Diminishing returns}

The sub-unity exponent for $N_{S}$ lends itself to widely-observed mathematical behaviors. Systems as seemingly
disparate as droplets \cite{PruppacherKlett1997}, boundary layers \cite{Turner1979}, animals \cite{Kooijman2010}
and plants \cite{Montieth2000} show growth behavior that is initially
rapid but slows with time, in what might be termed a {}``law
of diminishing returns''. 

To see why, consider that flows evolve
at rate $\eta=d\ln j/dt$ (Eq. \ref{eq:dlnadt}) where $j\propto N_{S}^{1/3}N_{R}$.
Thus, from Eq. \ref{eq:dNcdt_NcNe} \begin{eqnarray}
\eta & = & \frac{1}{3}\left(\frac{\partial\ln N_{S}}{\partial t}\right)_{N_{R}}+\left(\frac{\partial\ln N_{R}}{\partial t}\right)_{N_{S}}\label{eq:eta_NcNe}\\
 & = & \frac{1}{3}\eta_{S}-\eta_{R}\label{eq:eta_etac-etae}\end{eqnarray}
Here, $\eta_{S}$ and $\eta_{R}$ represent the respective growth
rates of the system and the reservoir, assuming the other is held
fixed. The rate $\eta_{S}$ represents the positive feedback that comes from system  
expansion. Growth lengthens the interface with respect to previously inaccessible reservoirs, allowing for increasing flows (Fig. \ref{fig:thermschematic}). The rate $\eta_{R}$ is a negative
feedback since reservoirs are simultaneously being depleted.

A Hamiltonian system with two potentials $\mu_{S}$ and $\mu_{R}$,
and no external sources to the volume $V$, is characterized by $N_{S}+N_{R}=N_{T}$
and $dN_{T}/dt=0$. Then, from Eq. \ref{eq:dNc_dNe}, Eq. \ref{eq:eta_NcNe}
can be rewritten as: \begin{eqnarray}
\eta & =\frac{d\ln{j}}{dt} = & j\left(\frac{1}{3N_{S}}-\frac{1}{N_{R}}\right)\label{eq:eta_etac-etae_j}\end{eqnarray}
which suggests a dimensionless {}``Adjustment number'' expressing
whether the evolution of flows is dominated by negative or positive
feedbacks: \begin{equation}
A=\frac{\eta_{S}}{3\eta_{R}}=\frac{N_{R}}{3N_{S}}\label{eq:A}\end{equation}
Flows are in a mode of either emergent growth or decay depending on
whether $A$ is greater or less than unity, respectively. 

Substituting Eq. \ref{eq:A} into Eq. \ref{eq:eta_etac-etae_j}, the
expression for the evolution of flows becomes \begin{equation}
\eta\left(t\right)=\alpha kN_{S}^{1/3}\left(A-1\right)\label{eq:growth_NT}\end{equation}
Decay dominates when $A<1$, in which case \begin{equation}
\eta\simeq\eta_{R}=\alpha kN_{S}^{1/3}\label{eq:decay}\end{equation}
Emergent growth requires that $A\gg1$, in which case $\eta\simeq\eta_{S}/3$,
where \begin{equation}
\eta_{S}=\alpha k\frac{N_{R}}{N_{S}^{2/3}}\label{eq:eta_growth}\end{equation}
Note that if it had been assumed that $j\propto N_{S}N_{R}$ rather
than $N_{S}^{1/3}N_{R}$, then emergent growth rates would have depended
only on the reservoir size $N_{R}$, and not on $N_{S}=\int_{0}^{t}jdt'$,
and therefore on past flows. Rather, as shown by Eq. \ref{eq:eta_growth},
growth rates have a power-law relationship given by $N_{S}^{-2/3}$, or the ratio of system
length and volume. 

The reason that growth in flows stagnates is that current flows are proportional to system
length $N_{S}^{1/3}$ (Eq. \ref{eq:dNs_jump}), while length grows
one third as fast as volume. Thus,  current flows
become progressively diluted in the volume accumulation of past flows
$N_{S}=\int_{0}^{t}jdt'$, and large systems tend to
grow at a slower rate and with lower thermodynamic efficiency
$\varepsilon=\eta/\alpha$ (Eq. \ref{eq:eta_growth}). 

Mathematically, if a system is in its emergent growth stage, such that $A\gg1$,
then its rate of growth evolves at rate\begin{equation}
\frac{d\ln\eta_{S}}{dt}\simeq-\frac{2}{3}\eta_{S}\label{eq:dlnetadt_growth}\end{equation}
While the system growth rate $\eta_{S}$ stays positive, its own
rate of change $d\ln\eta_{S}/dt$ is negative. The solution to
Eq. \ref{eq:dlnetadt_growth} is \begin{equation}
\eta_{S}\left(t\right)=\frac{\eta_{S0}}{1+2\eta_{S0}t/3}\label{eq:growth_soln}\end{equation}
where, $\eta_{S0}$ is the initial value of $\eta_{S}$ at time $t=0$.
Provided the system is initially small (i.e. $N_{S}\ll N_{R}$), its
growth rate has a half life of $3/\left(2\eta_{S}\right)$. Eq. \ref{eq:growth_soln}
accounts for the phenomenon of a {}``law of diminishing returns''
where a system is growing in response to conserved flows from a potential
energy reservoir. Relative growth rates start quickly, but they asymptote
to zero over time. Current flows become diluted in past flows
\footnote{As shown in the Appendix, $\eta_{S}$ is equivalent
to the local rate of entropy production.}.

\section{Logistic and explosive growth }

Two phenomena often seen in physical, biological and
social systems are sigmoidal growth \cite{Cohen1995,Tsoularis2002}
and super-exponential (sometimes termed {}``faster than exponential''),
or {}``explosive'' growth \cite{Bettencourt2007,GarrettCO2_2009}.
Sigmoidal behavior, as described by the logistic equation, starts
exponentially but saturates. By contrast, explosive instabilities exhibit rates
of change that grow super-exponentially with time, such that $\eta_{S}$ and $d\ln\eta_{S}/dt$
are both greater than zero. One immediately recognizable example is
the historically explosive growth of the world population \cite{Pollock1988,Johansen2001}. 

Explosive growth requires that the reservoir $N_R$ be open to some external source. Then, from Eq. \ref{eq:eta_growth}
the system growth rate $\eta_{S}$ evolves at rate\begin{equation}
\frac{d\ln\eta_{S}}{dt}=-\frac{2}{3}\eta_{S}+\eta_{R}^{net}\label{eq:dlnetaSdt}\end{equation}
where $\eta_{R}^{net}=d\ln N_{R}/dt=\eta_{D}-\eta_{R}$ represents
a balance between rates of reservoir discovery $\eta_{D}$ due to flows into the reservoir, and depletion $\eta_{R}$ due to flows out of the reservoir into the system. This suggests a {}``Growth Number'' \begin{equation}
G=\frac{3}{2}\frac{\eta_{R}^{net}}{\eta_{S}}\label{eq:G}\end{equation}
Explosive growth with $d\ln\eta_{S}/dt>0$ is possible provided that $G>1$, in which case the reservoir is growing at least two-thirds as fast as the system is growing. Steady-state
growth occurs when $G=1$ and $\eta_{R}^{net}=2\eta_{S}/3$. 
\begin{figure}
\includegraphics[width=7cm]{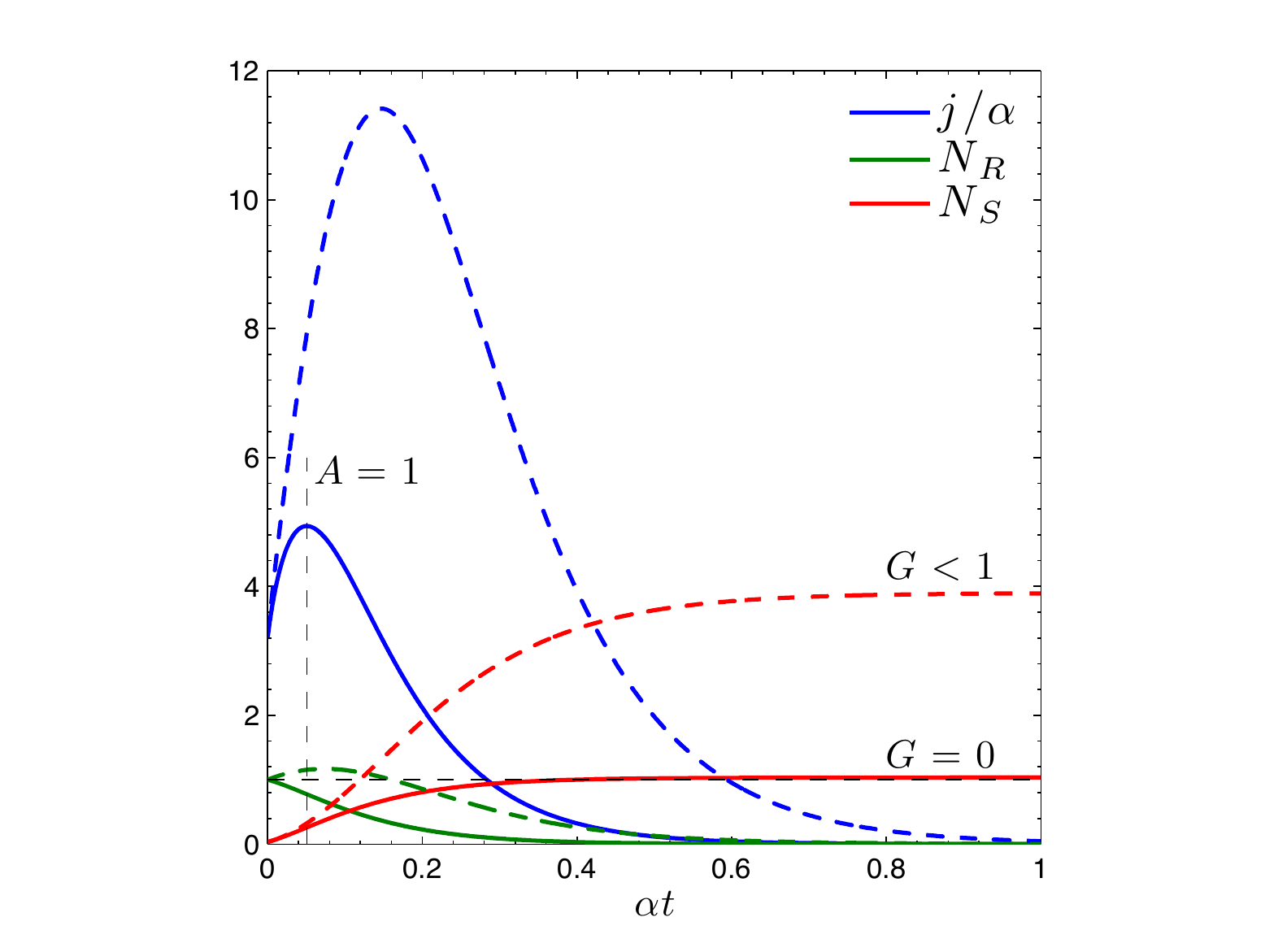}

\caption{\label{fig:NSNR}Numerical solution for the evolution of fluxes $j$
and the reservoir size $N_{R}$ and the system size $N_{S}$ for two
cases. Solid lines: $N_{S}+N_{R}$ is a constant and the initial value
of $G=0$ (Eq. \ref{eq:G}); peak flows occur where $A=N_{R}/\left(3N_{S}\right)=1$
(Eq. \ref{eq:A}). Dashed lines: there is {}``discovery'' of new
reservoirs $\eta_{D}>0$, but at a rate that is smaller than what
is required for super-exponential growth, so $G<1$. In both cases
reservoirs ultimately give way to net depletion (i.e. $\eta_{R}^{net}<0$).}

\end{figure}

\begin{figure}
\includegraphics[width=7cm]{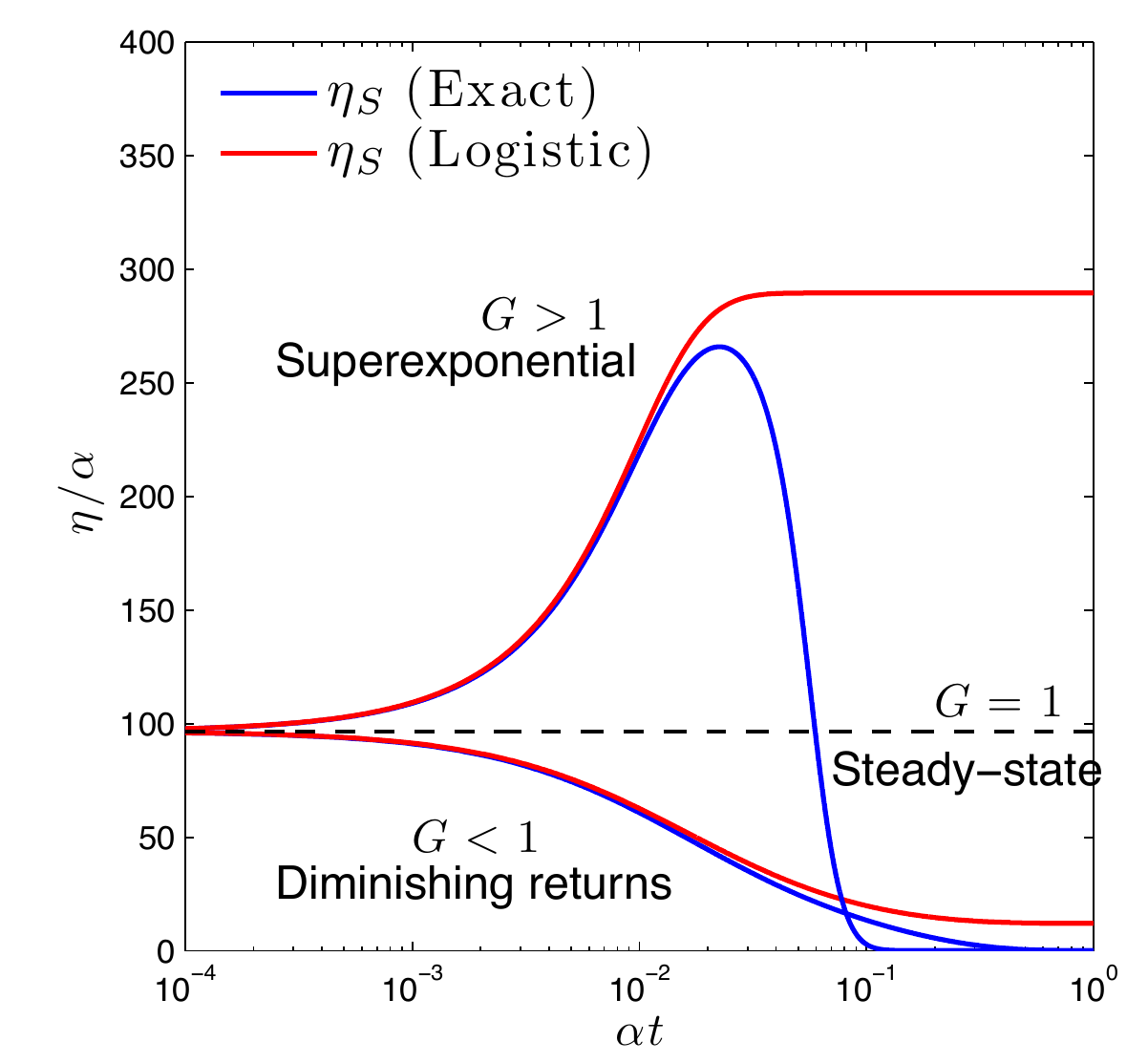}

\caption{\label{fig:Gplot}The growth rate of $\eta_{S}=d\ln N_{S}/dt$ as
a function of time for three regimes of the growth number $G$ (Eq.
\ref{eq:G}). Red lines show analytical solutions for emergent growth
given by the logistic expression in Eq. \ref{eq:soln}. Exact numerical
solutions given by the blue lines account for how flow rates eventually
decay as $N_{R}$ is depleted.}

\end{figure}

Eq. \ref{eq:dlnetaSdt} is expressible as a logistic equation for
rates of growth \begin{equation}
\frac{d\eta_{S}}{dt}=\eta_{R}^{net}\eta_{S}-\frac{2}{3}\eta_{S}^{2}\label{eq:slopeandintercept}\end{equation}
The prognostic solution for Eq. \ref{eq:slopeandintercept}, with
initial conditions given by $G=3\eta_{S0}/2\eta_{R0}^{net}$, is of
standard sigmoidal form \begin{equation}
\eta_{S}\left(t\right)=\frac{G\eta_{S0}}{1+\left(G-1\right)e^{-\eta_{R0}^{net}t}}\label{eq:soln}\end{equation}
The growth rate $\eta_{S}$ adjusts sigmoidally to $G\eta_{S0}$,
or 50\% faster than the net energy reservoir expansion rate $\eta_{R0}^{net}$.
Figures illustration the logistic nature of emergent growth rates,
and how they ultimately give way to reservoir depletion, are shown
in Figs. \ref{fig:NSNR} and \ref{fig:Gplot}.

\section{Rapid production of cloud droplets and rain}

One example of how instability can lead to runaway explosive growth
is in the formation of embryonic raindrops. The growth of the droplet
radius through vapor diffusion is constrained by a law of diminishing
returns. Production of embryonic raindrops requires a rapid transition
of cloud droplet size from about 10 $\mu$m to 50 $\mu$m radius through
interdroplet collision and coalescence \cite{Langmuir1948,PruppacherKlett1997}.
What remains poorly explained is how this {}``autoconversion'' process
can happen as rapidly as has been observed \cite{Wang2006}. 

Within the context of the discussion above, consider a droplet population
with number density $n_{d}$, each having volume $V_{S}$ and bulk
molecular density $n_{l}=N_{S}/V_{S}$. Eq. \ref{eq:decay} becomes
the relaxation rate of the available vapor supply in reponse to condensational
flows $\eta_{R}=4\pi r\mathcal{D}n_{d}$ \cite{Squires1952,Kostinski2009}.
Eq. \ref{eq:eta_growth} for system (or droplet volume) growth becomes
\begin{equation}
\eta_{S}=\frac{3\mathcal{D}\Delta n_{v}}{n_{l}r^{2}}\label{eq:eta_c_droplet}\end{equation}
where $\Delta n_{v}=n_{v}-n_{v}^{sat}=N_{R}/V$ is the local vapor
density surplus relative to the saturation value $n_{v}^{sat}$ at
the droplet surface \cite{BakerCorbinLathamI1980}. Note how growth
rates slow as $r$ grows. 

A droplet can overcome this law of diminishing returns by {}``discovering''
new mass reservoirs through the droplet collision-coalescence process.
If droplets are generally uniformly distributed and efficiently collected, with a dimensionless mass mixing ratio in air of $q_{l}$, then
a larger, falling, collector droplet with mass $m$ will grow through collisions at rate \begin{equation}
\eta_{D}=\frac{d\ln m}{dt}=Cq_{l}r\label{eq:etaD-droplet}\end{equation}
where $C\sim10^{5}m^{-1}s^{-1}$ (Details in Appendix).
If the depletion of droplets through this process $\eta_{R}$ remains small, then $\eta_{R}^{net}\simeq\eta_{D}$
and the collision-coalescence leads to explosive growth provided that
Eq. \ref{eq:G} satisfies \begin{equation}
G=\left(\frac{Cn_{l}}{2\mathcal{D}\Delta n_{v}}\right)^{1/3}q_{l}^{1/3}r>1\label{eq:rthresh}\end{equation}
For example, conditions characteristic of a small cumulus cloud might have a liquid
mixing ratio $q_{l}$ of 0.5 g kg$^{-1}$ and a supersaturation $S=\Delta n_{v}/n_{v}^{sat}$
of 0.5\%. In this case, explosive droplet growth could be theoretically expected
provided that a fraction of the droplet population exceeded a radius of about 20
$\mu$m. This is in fact  the threshold radius that is commonly observed as being necessary for warm rain production \cite{RangnoHobbs2005}.

\section{Thermodynamics of predator-prey relationships}

If, in addition to a source, a sink for a system is explicitly resolved in Fig. \ref{fig:thermschematic}, then the logistic expressions for $N_{S}$ and $N_{R}$ can be expressed in terms of predators and prey, as commonly considered in the ecological sciences and more recently
for physical representations of stratocumulus cloud dynamics \cite{Feingold2010,KorenFeingold2011}.
A fall in predators is followed by a rise in prey. The response is
renewed predation at the sacrifice of the prey. This oscillatory behavior
is canonically represented by the Lotka-Volterra equations \cite{Lotka1925},
which represent the one-way fluxes of populations of prey to predators
in terms of the product of the biomass densities of each, i.e, $N_{S}N_{R}$.
Many improvements to this model have been made over the past century
in order to more faithfully reproduce observed behavior, but not necessarily
by appealing to physical conservation laws \cite{Berryman1992}.

\begin{figure}
\includegraphics[width=6cm]{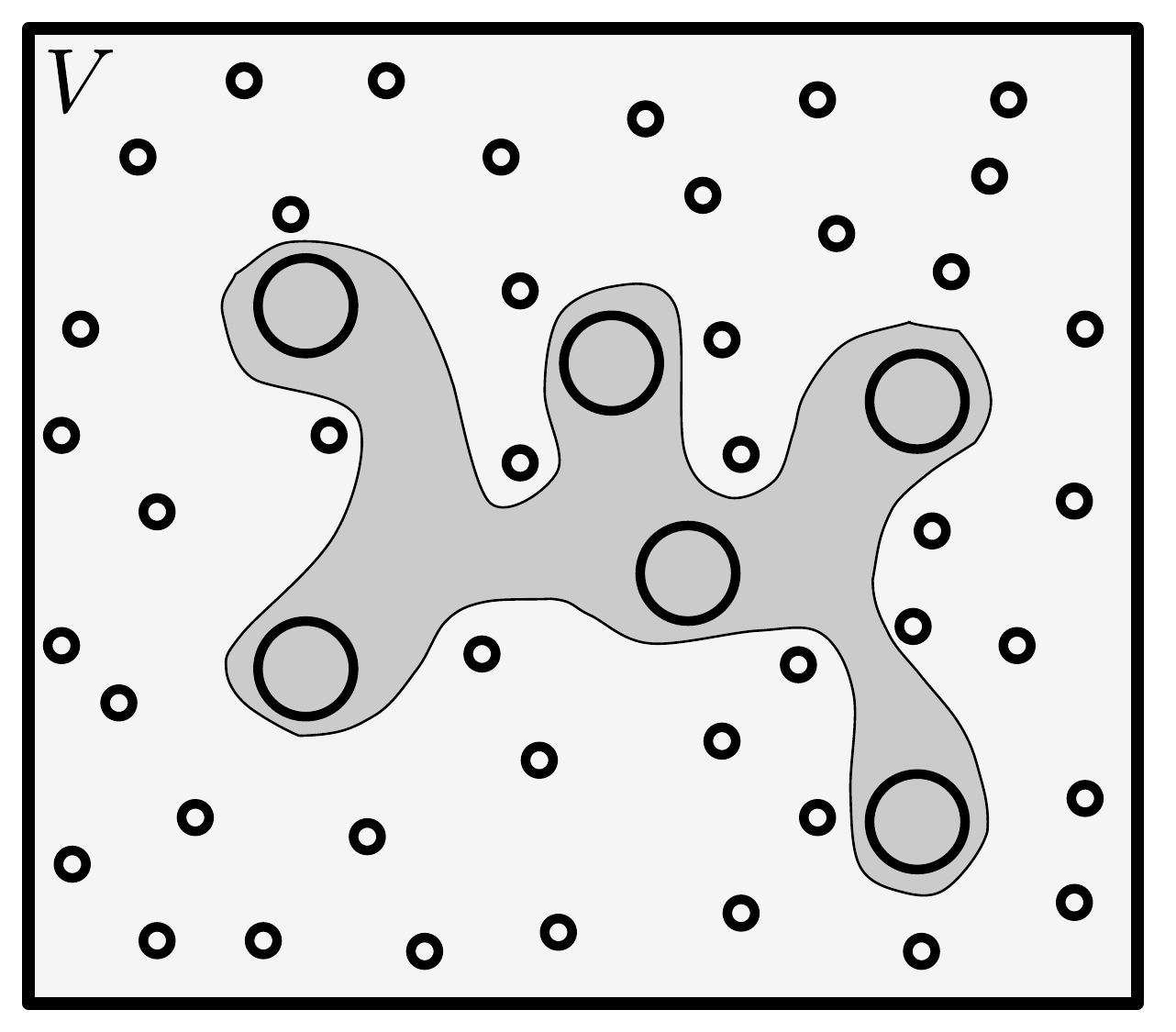}

\caption{\label{fig:Speciesinteractions} A predator population (large circles) acts as a system that interacts with a reservoir of prey (small circles) through an arbitrarily shaped linear interface within a total volume $V$.}

\end{figure}

The physical framework discussed here can be interpreted as a one-way
material flow of {}``prey'' biomass $N_{R}$ to {}``predator''
biomass $N_{S}$ . As discussed above, representing species interactions
as a product of predator and prey populations, e.g. $N_{S}N_{R}$, would seem to require the unphysical condition that predators and prey interact in the absence of a local
gradient. Physically, this is best addressed by introducing an arbitrarily shaped interface (Fig. \ref{fig:Speciesinteractions}), requiring that interactions be proportional to $N_{S}^{1/3}N_{R}$. In this case, the modified predator-prey relationships are \begin{eqnarray}
\frac{dN_{R}}{dt} & = & \beta N_{R}-\gamma N_{S}^{1/3}N_{R}\label{eq:LV}\\
\frac{dN_{S}}{dt} & = & \gamma N_{S}^{1/3}N_{R}-\delta N_{S}\nonumber \end{eqnarray}
where $\beta$, $\gamma$ and $\delta$ are constant coefficients.
The coefficient $\beta$ is equivalent to the discovery rate $\eta_{D}$
discussed previously, $\gamma=\alpha k$ (Eq. \ref{eq:dNcdt_NcNe}),
and $\delta$ represents the sink rate of $N_{S}$ to its surroundings,
as shown in Fig. \ref{fig:thermschematic}.

\begin{figure}
\includegraphics[width=7cm]{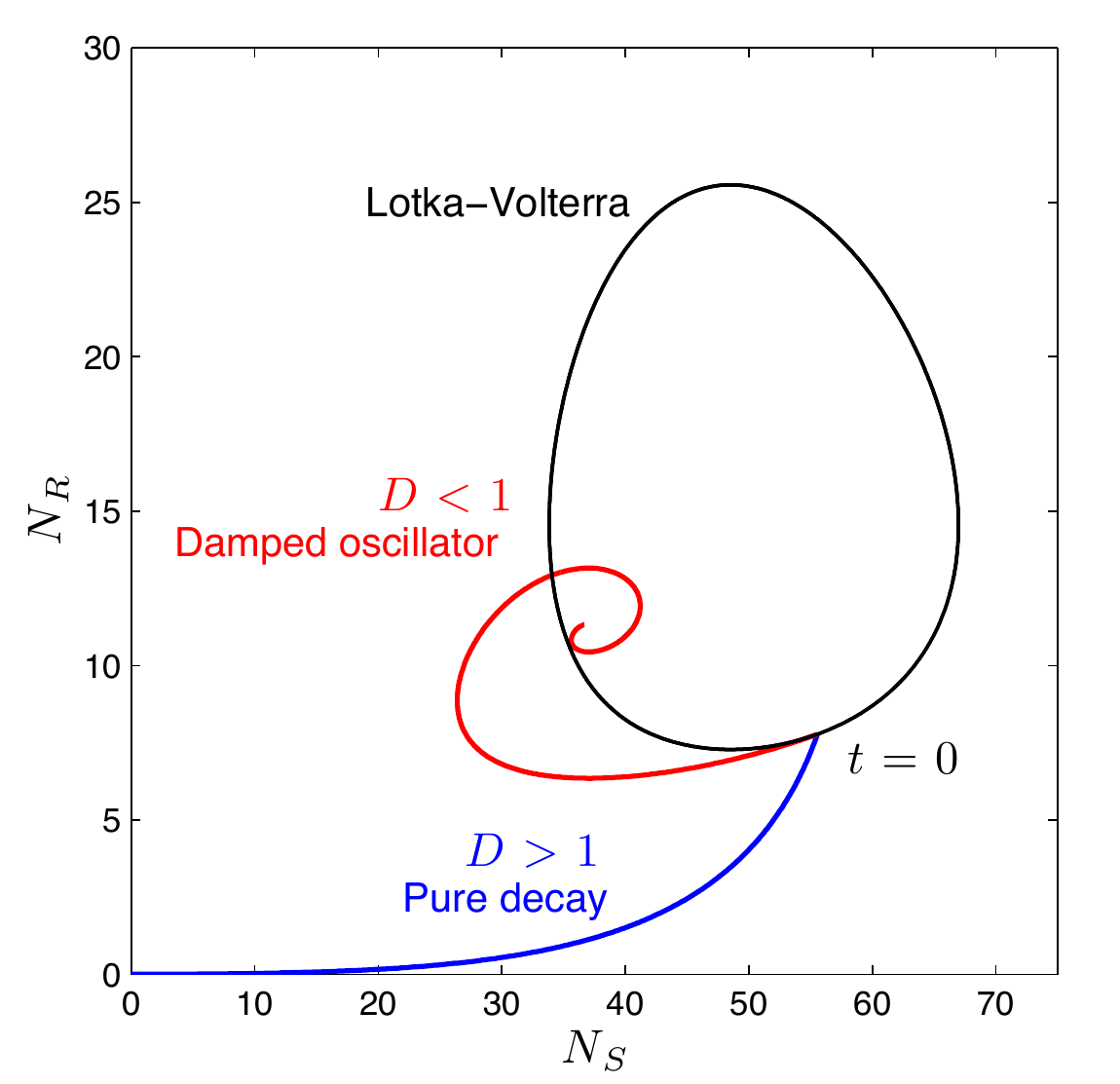}

\caption{\label{fig:LV}Phase plot for $N_{S}$ and $N_{R}$ for the predator-prey
equations given by Eq. \ref{eq:LV} (blue and red), depending on the
value of the damping number $D=\delta/\left(3\beta\right)$. The limit
cycle behavior given by the canonical Lotka-Volterra Equations (black)
is shown for the same set of initial conditions.}

\end{figure}

So, while the Lotka-Volterra equations lead to non-dissipative limit
cycles, the simple addition of a one third exponent to the predators allows
populations to converge on an equilibrium state given by $N_{R}=\delta\beta^{2}/\gamma^{3}$
and $N_{S}=\left(\beta/\gamma\right)^{3}$, or $N_{R}/N_{S}=\delta/\beta$.
As shown in Appendix and in Fig. \ref{fig:LV},
the nature of convergence depends on a damping number $D=\delta/\left(3\beta\right)$.
If $D<1$, then $N_{S}$ behaves as a damped simple harmonic oscillator
with angular frequency $\omega_{1}=\omega_{0}\sqrt{1-D}$, where $\omega_{0}=\left(\delta\beta/3\right)^{1/2}$.
If $D\geq1$, then equilibrium is approached in monotonic decay. 

The key point here is that the sub-unity exponent for $N_{S}$ allows for inter-species interactions to evolve more slowly than the respective populations themselves, introducing
a damped or {}``buffered'' \cite{KorenFeingold2011} response. The general perturbation solution for the damping of  $N_{S}$ is $a\exp\left(-\delta t\right)\exp\left[i\left(\omega_{1}t-\phi\right)\right]$ where $a$ and $\phi$ are determined by the initial conditions.

Damped simple harmonic oscillators are ubiquitous in physics, for
example in the interactions of light with particles \cite{Liou-book}, 
so it is particularly noteworthy that it requires only a very small modification to the Lotka-Volterra predator-prey
framework in order to arrive at an expression of this form. In fact, even in ecology,
damped oscillatory behavior is being observed in the response of
forage fish populations to a collapse of predatory cod stocks from
overfishing \cite{FrankPetrie2011}. From the above, a possible interpretation
is that forage fish biomass densities ($N_{S}$) initially thrived
when predator cod stocks collapsed (a drop in $\delta$), but then
they overshot and declined themselves as a consequence of excessive
plankton ($N_{R}$) depletion at rate $\gamma N_{S}^{1/3}N_{R}$.
As forage fish populations fell, plankton recovered at rate $\beta N_{R}$,
and the forage fish soon followed. Equilibrium is being restored, and in the manner of damped oscillations. 

A more physical parameter space for the phase diagram shown in Fig.
\ref{fig:LV} could be either the orthogonal basis of $j=\gamma N_{S}^{1/3}N_{R}$
and $dN_{S}/dt$, or alternatively $N_{S}^{1/3}$ and $N_{R}$. For
the latter, the area carved out by the curves in a phase diagram would
be proportional to the energy dissipated by the trophic cascade.

\section{Summary}

Regardless of scale or complexity, anything that can be defined requires
some local contrast to be observable. Contrasts require a gradient
and therefore a local exchange of material and energetic flows. Physically,
flows are across an interface that is related to the magnitude of
the local gradient, normal to the surface of the system. Dimensional reasoning requires that flows must be proportional to
a length dimension, or a one third exponent with respect to the system
volume or mass. 
\begin{table}[h]
\caption{\label{tab:Evolutiontable}Evolutionary modes for emergent system
growth rates. }
\begin{tabular}{cccc}
\hline 
 & $d\ln\eta_{S}/dt$ & $\eta_{R}^{net}$  & $G$\tabularnewline
\hline
explosive growth & $>0$ & $>2\eta_{S}/3$ & $>1$\tabularnewline
 steady state & $0$ & $=2\eta_{S}/3$ & $1$\tabularnewline
diminishing returns & $<0$ & $<2\eta_{S}/3$ & $<1$\tabularnewline
\hline
\end{tabular}
\end{table}

The consequence of the one third exponent is that the time evolution of flow rates follows mathematical
behaviors that can be partitioned into a limited set of regimes (Table
\ref{tab:Evolutiontable}). In general, spontaneous emergence is
governed by the logistic equation, exhibiting a sigmoidal curve for
system growth rates. The one third exponent requires that current
flows become increasingly diluted in an accumulation of past flows,
so spontaneously emergent systems have a natural propensity to exhibit
a law of diminishing returns. Explosive, faster-than-exponential growth
occurs if energy reservoirs are expanding at least two-thirds as fast
as the rate of system growth. However, even explosive growth ultimately
lends itself towards decay in flow rates. The faster a system grows,
the faster it depletes its potential energy reservoirs.

Where a system is open to downhill flows to and from it, the system
size itself can either grow or decay, depending on the sign of net convergence
in flows. In this case, the growth equations are very similar to the
canonical Lotka-Volterra predator-prey equations used to model ecological
systems, differing only in a one third exponent. This subtle but important
difference leads to the perturbation equations for a damped simple
harmonic oscillator that are are ubiquitous in the physical sciences
and have also been identified in ecological systems. Whether
the oscillator is under- or over-damped depends on the ratio of the
natural growth rates for {}``predators'' and {}``prey''.

The mathematical expressions described here are independent of scale
or complexity, and any physics more specific than thermodynamic laws.
They offer a simple framework for expressing how a redistribution of 
matter and energy evolves through a cascading flow between distinguishable
systems.

\begin{ack}{This work was supported by the Kauffman Foundation}
 \end{ack}
 
 \begin{appendix}{}
\section{Appendix}
 \subsection{Entropy production}

The equation for the growth rate of $N_{S}$ is

\begin{eqnarray}
\eta_{S} & = & \frac{d\ln N_{S}}{dt}=\frac{j}{N_{S}}\label{eq:eta_etac-etae_j-SI}\end{eqnarray}
$N_{S}$ is an accumulation of past flows, so this can be rewritten
as \[
\eta_{S}=\frac{j}{\int_{0}^{t}jdt'}\]
The current growth rate of flows $\eta_{S}$ is tied to the integrated
history of past flows $\int_{0}^{t}jdt'$. 

The potential energy dissipation rate $a$ along the gradient $\Delta\mu$
is proportional to the material flow $j$ down the gradient.
Thus

\begin{eqnarray*}
\eta_{S} & = & \frac{a}{\int_{0}^{t}adt'}\end{eqnarray*}
The expression $\int_{0}^{t}adt'$ is the total time-integrated heating
that has been applied to the constant potential surface $\mu_{S}$.
The local rate of production of entropy $\mathcal{S}$ can be written
as the energy dissipation rate relative to the local potential, i.e.
$\sigma=d\mathcal{S}/dt=a/\mu_{S}$. It follows that the accumulation
of entropy within the volume $V$ that contains fixed potentials $\mu_{R}$
and $\mu_{S}$ is $\mathcal{S}=\int_{0}^{t}adt'/\mu_{S}$. Thus, $\eta_{S}$
has the thermodynamic expressions

\[
\eta_{S}=\frac{d\ln N_{S}}{dt}=\frac{\sigma}{\mathcal{S}}=\frac{d\ln\mathcal{S}}{dt}=\frac{a}{\int_{0}^{t}adt'}\]
Energy dissipation at rate $a$ drives conservative material flows
at rate $j$ from a high potential $\mu_{R}$ to a lower potential
$\mu_{S}$. The growth rate of the amount of material in the lower
potential is proportional to the rate at which entropy is increasing
locally through $\sigma=\mathcal{S}d\ln N_{S}/dt$.

\subsection{Collision-coalescence}

The growth equation for the mass $m=4\pi\rho_{l}r^{3}/3$ of a collector
drop with radius $r$ and density $\rho_{l}$, that falls with terminal
velocity $v_{T}$ through a cloud of droplets with liquid water mixing
ratio $q_{l}$ is \[
\frac{dm}{dt}\simeq\pi r^{2}v_{T}\rho_{air}q_{l}\]
where $\rho_{air}$ is the air density, and it is assumed that the
collector drop has a relatively large cross-section and the collection
efficiency is near unity. In the initial stages of growth, when the
collector drop is smaller than about 35 $\mu$m, the drop terminal
velocity is determined by a balance between Stokes drag and the gravitational
force $mg$, such that \[v_{T}=\frac{2\rho_{l}g}{9\rho_{air}\nu}r^{2}\]
where $\nu$ is the kinematic viscosity of air. Thus,\[
\eta_{D}=\frac{d\ln m}{dt}=\frac{g}{6\nu}q_{l}r\simeq Cq_{l}r\]
where $C\simeq10^{5}$~m$^{-1}$~s$^{-1}$.

\subsection{Perturbation solutions for the predator-prey equations}

The original set of predator-prey equations is 

\begin{eqnarray}
\frac{dN_{R}}{dt} & = & \beta N_{R}-\gamma N_{S}^{1/3}N_{R}\nonumber \\
\frac{dN_{S}}{dt} & = & \gamma N_{S}^{1/3}N_{R}-\delta N_{S}\label{eq:NR_NS}\end{eqnarray}
 which can be re-written in a more amenable mathematical form as \begin{eqnarray}
\frac{dx}{dt} & = & \beta x-\gamma xy\nonumber \\
3y^{2}\frac{dy}{dt} & = & \gamma xy-\delta y^{3}\label{eq:xy}\end{eqnarray}
 where $x=N_{R}$ and $y=N_{S}^{1/3}$. The equilibrium solutions
for $x$ and $y$ are $x_{eq}=\delta\beta^{2}/\gamma^{3}$ and $y_{eq}=\beta/\gamma$. 

Supposing a perturbation solution \begin{eqnarray}
x & = & \frac{\delta\beta^{2}}{\gamma^{3}}+x'\label{eq:xy_perturbation}\\
y & = & \frac{\beta}{\gamma}+y'\nonumber \end{eqnarray}
and noting that $dx'/dt=dx/dt$ and $dy'/dt=dy/dt$, Eqs. \ref{eq:xy}
are transformed to \begin{eqnarray}
\frac{dx'}{dt} & = & -\left(\frac{\delta\beta^{2}}{\gamma^{2}}\right)y'\label{eq:dxydt_perturbations}\\
\frac{dy'}{dt} & = & \left(\frac{\gamma^{2}}{3\beta}\right)x'-\left(\frac{2\delta}{3}\right)y'\nonumber \end{eqnarray}
where second order perturbation terms have been neglected. Taking
the second derivative leads to the equation for a damped simple harmonic
oscillator \begin{equation}
\frac{d^{2}y'}{dt^{2}}+\frac{2\delta}{3}\frac{dy'}{dt}+\frac{\delta\beta}{3}y'=0\label{eq:wave_equation}\end{equation}
The natural oscillator angular frequency is $\omega_{0}=\left(\delta\beta/3\right)^{1/2}$.
Eq. \ref{eq:wave_equation} has the general solution $y'=a\exp^{\eta_{y1}t}+b\exp^{\eta_{y2}t}$
where $\eta_{y1}$ and $\eta_{y2}$ are the quadratic roots \begin{equation}
\eta_{y}=\frac{\delta}{3}\left[-1\pm\sqrt{1-3\frac{\beta}{\delta}}\right]\label{eq:p}\end{equation}
Since the real part of $\eta_{y}$ is always negative, $y'$ always
decays. The nature of the decay depends on a damping ratio \begin{equation}
D=\frac{\delta}{3\beta}\label{eq:Dappend}\end{equation}
The value of $\eta_{y}$ is complex if $D<1$, in which case decay
is oscillatory with frequency\[
\omega_{1}=\omega_{0}\sqrt{1-D}\]
In terms of $N_{S}$, for the real component $\eta_{S}=3\eta_{y}$
since $d\ln y/dt=\left(d\ln N_{S}/dt\right)/3$. Thus, the solution
for $N_{S}$ is\[
N_{S}=a\exp\left(-\delta t\right)\exp\left[i\left(\omega_{1}t-\phi\right)\right]\]
 where $a$ and $\phi$ are determined by the initial conditions.

\begin{figure}
\includegraphics[width=8.7cm]{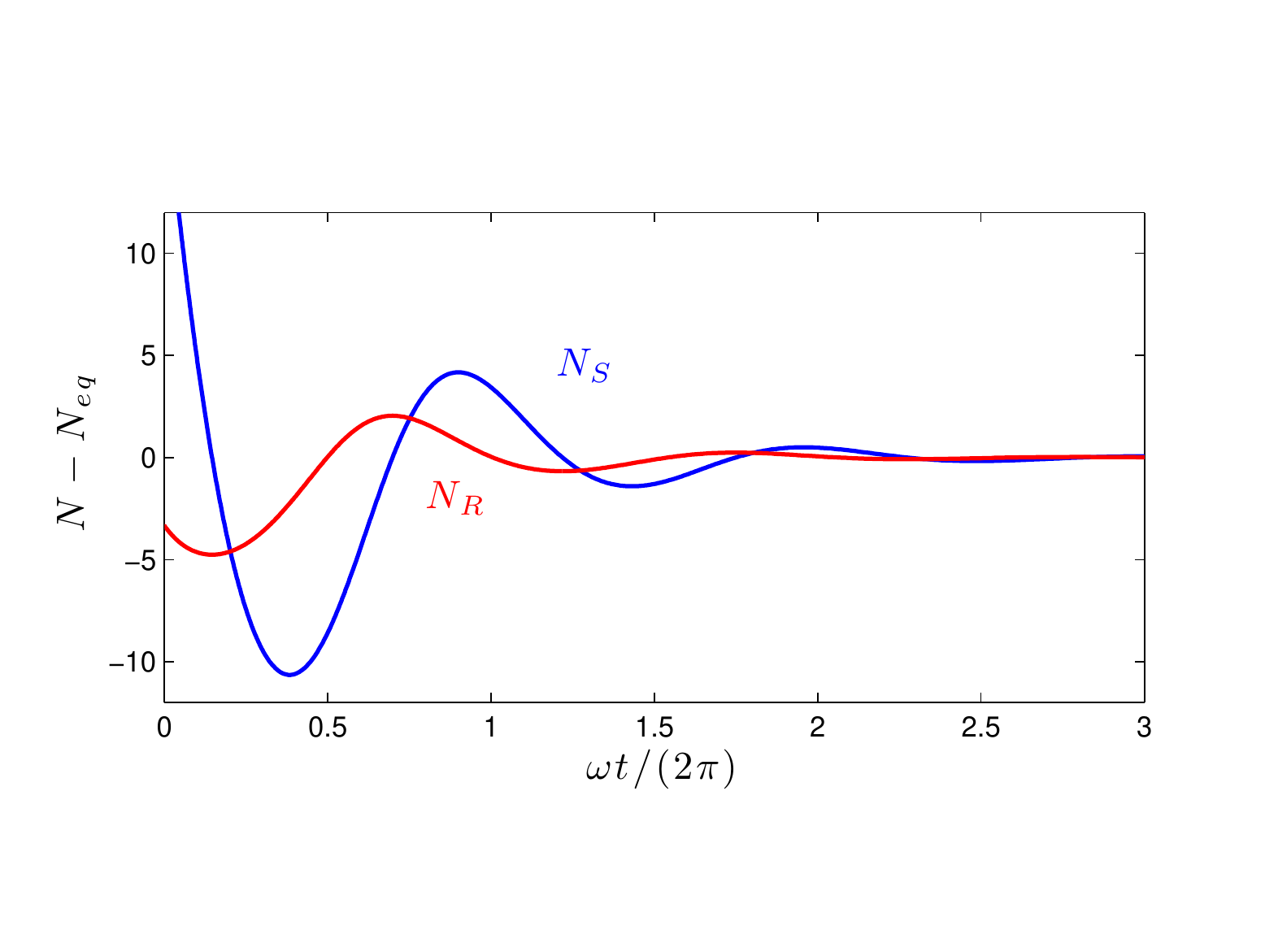}

\caption{\label{fig:LV-1}Time series for perturbation from equilibrium in
$N_{S}$ and $N_{R}$ for the predator-prey equations given by Eq.
\ref{eq:NR_NS}, for the special case that $D<1$ and the perturbations
behave as damped simple harmonic oscillators.}
\end{figure}

\end{appendix}

\end{document}